\documentclass[pre]{revtex4}
\usepackage{amssymb,bm}
\usepackage[dvips]{graphicx}
\usepackage{psfrag}

\begin{document}

\title{Spatial statistics of magnetic field in two-dimensional chaotic flow in the resistive growth stage.}

\author{I. V. Kolokolov}

\affiliation{Landau Institute for Theoretical Physics RAS, \\
 119334, Kosygina 2, Moscow, Russia, }
 \affiliation{NRU Higher School of Economics, \\
101000, Myasnitskaya 20, Moscow, Russia.}

\date{\today}

\begin{abstract}

The correlation tensors of magnetic field in a two-dimensional chaotic flow of conducting fluid are studied.
It is shown that there is a stage of resistive evolution where the field correlators grow exponentially with time what contradicts to the 
statements present in literature. The two- and four-point field correlation tensors are computed explicitly in this stage in the framework of Batchelor-Kraichnan-Kazantsev model. 
 These tensors demonstrate highly 
intermittent statistics of the field fluctuations both in space and time. 

\end{abstract}


\maketitle

\section{Introduction}

Kinematic dynamo consists in enhancement of magnetic field fluctuations in a moving conducting fluid. This enhancement is statistically significant for non-stationary flows, 
in particular, chaotic flows. The dynamo is 	 
relevant for astrophysics  \cite{MHD,P79,ZRS,astro1,schek} and for so-called elastic turbulence which 
is the chaotic motion of polymer solutions \cite{00GS,01GSb,00BFL}. In both cases the velocity field $\bm{v}(\bm{r},t)$ can be considered 
as a smooth function of coordinates $\bm{r}$   allowing the use of the Taylor expansion. Practically this means that in the Lagrangian frame comoving with a given liquid particle the 
profile of the field $\bm{v}(\bm{r},t)$ is approximately linear:
 \begin{equation}
 v_\alpha(\bm{r},t)\approx \sigma_{\alpha\beta}(t)r_\beta
 \label{linprof}
 \end{equation}
with matrix-valued random function $\sigma_{\alpha\beta}(t)$ of time $t$. 
Evolution of the magnetic field $\bm{B}(\bm{r},t)$ is governed by the equation \cite{LL}:
 \begin{equation}
 \partial_t{\bm B}=
 ({\bm B}\cdot{\bm\nabla}){\bm v}
 -({\bm v}\cdot{\bm\nabla}){\bm B}
 + \kappa\nabla^2{\bm B},
 \label{magnet}
 \end{equation}
where $\kappa$ is the dissipation coefficient inverse proportional to the conductivity of the fluid. 

The present paper is devoted to  computation of  correlation tensors of the field ${\bm B}$ 
for two-dimensional incompressible chaotic flow at conditions when the expansion (\ref{linprof}) is applicable. The statistics of the traceless matrix 
$\sigma_{\alpha\beta}(t)$ is assumed to be the Gaussian $\delta$-correlated in time random process:
\begin{equation}
\langle \sigma_{\alpha\mu}(t)\sigma_{\beta\nu}(t')\rangle=
{\cal G}\left(3\delta_{\alpha\beta}\delta_{\mu\nu}-\delta_{\beta\nu}\delta_{\alpha\mu}-\delta_{\beta\mu}\delta_{\alpha\nu}\right)
\delta(t-t'),
 \label{sta}
 \end{equation}
where the Stratonovich regularization of the $\delta$-function is assumed,i.e.,  the $\delta$-function is treated as the limit of a sequence of symmetric with respect 
to reflection $(t-t')\to (t'-t)$ functions. It should be stressed that  this is an
external requirement and  important   part  of the model formulation.   
Such tensor structure follows from the expansion of the velocity field  $\bm{v}(\bm{r},t)$ correlator in Eulerian frame:  
\begin{equation}
\langle v_{\mu}(\bm{r},t)v_{\nu}(\bm{0},t')\rangle \approx \left[V_0^2\delta_{\mu\nu}-
{\cal G}\left(\frac{3}{2}r^2\delta_{\mu\nu}-r_\mu r_\nu\right)\right]
\delta(t-t'), \quad r\ll R_c,
 \label{velt}
 \end{equation}
where $R_c$ has a meaning of the smoothness scale determined by fluid viscosity. We consider here the case when $R_c\gg r_d=2\sqrt{\kappa/{\cal G}}$ which is suitable for astrophysical and 
rheological applications.

It was shown \cite{Kaz,99CFKV,schek} for three-dimensional chaotic flows described by this BKK (Batchelor-Kraichnan- Kazantsev) model that the  
moments $\langle \bm{B}^{2n}(\bm{r},t)\rangle, n=1,2, \dots$  grow exponentially with time and the correlation functions of the field tend to universal  spatial shapes.  
These shapes correspond to filaments with increasing length in which the magnetic field is concentrating in the course of evolution. The width of such filaments 
is decreasing up to the dissipative scale 
$r_d$ where the value if the width is stabilized. At this moment the resistive stage of the evolution begins. The exponential growth continues but the exponent changes
(see also review   \cite{01FGV}). 

Two-dimensional flows were not considered long time in dynamo studies since it was asserted in 
\cite{Zeldovich1, Zeldovich2,Zeldovich3} (and recognized by the community) that in this case the initial growth of the field changes to exponential decay when the evolution 
turns to be strictly diffusive. It should be emphasized that  the magnetic field ${\bm B}(\bm{r},t)$ is considered to be three-dimensional both as a vector and as a function of space position,
only the flow is two-dimensional: ${\bm v}=(v_1, v_2, 0)$. 
In the papers \cite{KogKolLe,KleSiz} it is shown explicitly that the arguments leading to this statement are incorrect and the non-dissipative initial growth
gives way to the dissipative regime which occurs with exponential increase of the magnetic field correlators again. When the characteristic length  of the filaments reaches $R_c$ 
 the field values in neighboring filaments begin to anticorrelate \cite{Kolo16} and the dynamo stops.
It is  in agreement with exact mathematical theorems \cite{Ose,Arno} related to infinite time limit. In \cite{Kolo16} it is shown that the maximal value of $\langle {\bm B}^2\rangle$ 
reached in the course of evolution is 
$\sim \langle {\bm B_0}^2\rangle (R_c/r_d)^2 \gg \langle {\bm B_0}^2\rangle$ where ${\bm B}_0$ is the initial field amplitude. This enhancement is picked up in the exponential growth stage.
Depending on initial level of fluctuations the growth stage can be terminated also by the back reaction of the magnetic field on the fluid flow \cite{schek,Sche2}. 
In this case the spatial structure of the field fluctuations is extremely important. We show in the present paper that the spatial statistics of the magnetic field in the resistive growth 
regime in the framework of  two-dimensional BKK-model is intermittent similar to three-dimensional case  \cite{99CFKV,01FGV}.  The path-integral formalism created to study passive scalar 
statistics \cite{ChGK,CFK} and isotropic correlations in three-dimensional dynamo problem \cite{99CFKV}  is  developed here to compute the tensor structure of two- and four-point 
magnetic field correlators.

The evolution equation (\ref{magnet}) conserves the divergence of the field  $\bm{B}(\bm{r},t)$. 
On the other hand, the evolution equations for the magnetic field components in the flow plane $B_\mu(\bm{r},t), \, \mu=1,2$ and for the 
component $B_3(\bm{r},t)$ perpendicular to this plane decouple. This means that we can consider evolution of two-component divergence-full field 
$B_\mu(\bm{r},t), \, \mu=1,2$ separately since the condition
\begin{equation}
\partial_\mu B_\mu+\partial_3 B_3=0
 \label{divergenziya}
 \end{equation}
is satisfied by the component $B_3(\bm{r},t)$ governed be the decoupled evolution equation. The dependence of $B_\mu(\bm{r},t), \, \mu=1,2$ 
on the coordinate $r_3$,  being essential in the condition (\ref{divergenziya}), is unimportant in the evolution. The coordinate $r_3$ enters 
the correlation functions of components $B_\mu(\bm{r},t), \, \mu=1,2$ as a parameter only and we don't take care here on it. 
 
\section{Dynamical computation of the  correlation tensors of the field $B_\alpha(\bm{r},t)$.} 
 
 The change of the frame to  (\ref{linprof}) does not change the simultaneous statistics  of the field $B_\mu(\bm{r},t)$, 
so that we use the following evolution equation:
 \begin{equation}
 \partial_t B_\alpha=
 \sigma_{\alpha\mu}B_\mu-\sigma_{\mu\nu}r_\nu\partial_\mu B_\alpha+
  \kappa\nabla^2 B_\alpha.
 \label{magnet1}
 \end{equation}
Performing the spatial Fourier transform:
\begin{equation}
 B_\alpha(\bm{r},t)=\int\frac{d^2{\bf k}}{(2\pi)^2}e^{i{\bf k}{\bf r}}b_\alpha(\bm{k},t)
 \label{fur}
 \end{equation}
we obtain for    $b_\alpha(\bm{k},t)$ the first-order partial differential equation:
 \begin{equation}
 \partial_t b_\alpha=
 \sigma_{\alpha\mu}b_\mu+\sigma_{\mu\nu}k_\mu\frac{\partial}{\partial k_\nu} b_\alpha-
  \kappa k^2 b_\alpha.
 \label{magnet2}
 \end{equation}
Its solution with the initial data $\bm{b}(\bm{k},0)$ has the form:
\begin{eqnarray}
b_\alpha\left( {\bm k},t\right) = (\hat{W})_{\alpha\beta}\left( t\right)b_\beta
\left( \hat{W}^T\left( t\right) {\bf k},0\right) \exp \left[ -\kappa
\int\limits_0^td\tau\left( {\bf k}\hat{W}(t,\tau)\hat{W}^T(t,\tau){\bf k}
\right) \right].
 \label{polle}
\end{eqnarray}
where the matrices $\hat{W}(t)$ and $\hat{W}(t,\tau)$ obey the equation 
\begin{equation}
d\hat{W}/dt=\hat{\sigma}\hat{W}
\label{weq}
\end{equation}
and can be written
as the ordered exponentials:
\begin{eqnarray}
\hat{W}(t)={\rm T}\exp \left( \int\limits_0^tdt' \,\hat{\sigma}\left( t'\right) \right), \quad  
\hat{W}(t,\tau)={\rm T}\exp \left(
\int\limits_{\tau}^t dt' \,\hat{\sigma}\left( t' \right) \right)= \hat{W}(t)\hat{W}^{-1}(\tau).
\label{upe}
\end{eqnarray}
The incompressibility of the flow $\sigma_{\alpha\alpha}=0$ leads to the unimodularity of the matrix $\hat{W}(t)$: $\det\hat{W}(t)=1$.

To perform the averaging over the matrix Gaussian random process $\hat{\sigma}(t)$ we use the path integral formalism. 
The  measure  corresponding to the correlation function (\ref{sta}) has the form: 
\begin{eqnarray}
{\cal D}\hat{\sigma}(\tau)\exp\left\{-\frac{1}{16 {\cal G}}\int \,d\tau 
\left[3\mathrm{Tr}\left(\hat{\sigma}\hat{\sigma}^T\right)+
\mathrm{Tr}\left(\hat{\sigma}^2\right)\right]\right\}.
 \label{ves}
 \end{eqnarray}
The matrix $\hat{W}(t)$ cannot be expressed as a functional of $\hat{\sigma}(t)$ explicitly. Hovewer, for the our problem 
this difficulty can be avoided (see also \cite{CFK}). Let us perform  the Iwasava parametrization of the matrix  $\hat{W}(t)$:
 \begin{equation}\label{tri}
\hat{W}=\hat{{\cal O}}(\varphi) \hat{D}(\rho)\hat{T}(\chi),  
\end{equation}
where 
\begin{equation}\label{tri1}
\hat{{\cal O}}(\varphi)= \left(\begin{array}{cc} \cos\varphi & \sin\varphi \\
        -\sin\varphi  & \cos\varphi  \end{array}\right),\quad
 \hat{D}(\rho)=
\left(\begin{array}{cc} e^{\rho} & 0 \\
        0  & e^{-\rho}  \end{array}\right),\quad
\hat{T}(\chi)=        
\left(\begin{array}{cc} 1 & \chi(t) \\
        0  & 1  \end{array}\right),
\end{equation}
and the parameters $\varphi(t), \rho(t)$ and $\chi(t)$ are some functions of time $t$ determined by $\hat{\sigma}(t)$ implicitly via the equation:
 \begin{equation}
 \label{tri2}
\hat{\sigma}(t)=d \hat{W}(t)/dt \hat{W}^{-1}(t)=\hat{{\cal O}}(\varphi)
\left(\begin{array}{cc} \dot{\rho} & \dot{\varphi}+\dot{\chi}e^{2\rho}\\
        -\dot{\varphi}  & -\dot{\rho}  \end{array}\right)\hat{{\cal O}}^{-1}(\varphi).
\end{equation}
The  initial conditions $\rho(0)=\chi(0)=\varphi(0)=0$ correspond to the evident equality $\hat{W}(0)=1$. If we consider the relation 
(\ref{tri2}) as the change of variables in the measure (\ref{ves}) we obtain an explicit path integral expression for any correlation functions of the magnetic field $B_\alpha(\bm{r},t)$. Assuming the retarded regularization of the derivatives, for example, 
$\dot{\rho}=(\rho_n-\rho_{n-1})/\epsilon$ where $\epsilon$ is the infinitesimal time interval, the Jacobian of the transformation can be set to  a constant.
The resulting averaging measure expressed in terms 
of the new variables has the form:
 \begin{eqnarray}
\label{mts}
{\cal N}
{\cal D}\chi{\cal D}\rho {\cal D}\varphi
\left(\prod_\tau e^{2\rho(\tau)}\right)
\exp\left\{-\frac{1}{2{\cal G}}\int\limits_0^t\! d\tau\! 
\left[\left(\dot{\rho}-{\cal G}\right)^2+\frac{1}{4}e^{4\rho}\dot{\chi}^2+
\frac{1}{2}\left(\dot{\varphi}+\dot{\chi}e^{2\rho}\right)^2 \right]\right\}.
\end{eqnarray}
Here the retarded regularization of the time derivatives is assumed (see e.g. \cite{Kam}) and ${\cal N}$ is the normalization  constant  providing the equality $<1>=1$. The origing of non-zero $\langle \dot{\rho}\rangle$ is due to 
so called contact terms in the Lagrangian like 
$\epsilon \dot{\chi}^2\dot{\rho}\exp(4\rho)\sim\dot{\rho}$ arising when we uniformize the regularization of all the terms  
to retarded form.   The corresponding explicit computations are simple but tedious. Instead we can fix the final coefficients by the requirement that
the growth law of a vector $\bm{a}(t)$ lenght:
\begin{equation}
 \langle \bm{a}^2(t)\rangle= \langle \bm{a}(0)\hat{W}^T(t)\hat{W}(t)\bm{a}(0)\rangle=\bm{a}^2(0)e^{4{\cal G}t}
 \label{vg}
 \end{equation}
following directly from (\ref{sta})  must be reproduced. It is worth noting that it is  equivalent to the equality 
$\langle \exp(2\rho) \rangle=\exp(4{\cal G}t)$ check.

We suppose the initial 
distribution of the magnetic field to be random with  Gaussian  spatially homogeneous statistics (below we consider also the case 
of a given configuration of the field). The correlator of the Fourier components of the field $b_\alpha(\bm{k},t)$ has the form:
\begin{equation}
\langle b_\alpha\left( {\bm k},0\right) b_\beta\left( {\bm k_1},0\right)=\delta\left({\bm k}+{\bm k_1}\right)
\delta_{\alpha\beta}f(k^2l^2),
 \label{incorr}
\end{equation} 
where $l$ is the  correlation lengh of the initial magnetic field fluctuations and $f(k^2l^2)$ is a positive function going  to zero when $kl\to \infty$. 
The expression for the correlation tensor of in-plane components of the magnetic field following from (\ref{polle}) and (\ref{incorr}) has the form:
\begin{eqnarray}
&F_{\alpha\beta}({\bf r},t)=\langle B_\alpha({\bf r},t) B_\beta(0,t)\rangle =
 \Bigl\langle\left[ \hat{W}(t)\hat{W}^{T}(t)\right]_{\alpha\beta}\times
\nonumber \\
&\times
\int \frac{d^2{\bf k}}{(2\pi)^2}
 f(k^2l^2)\exp\left(i{\bf r}\hat{W}^{-1}(t){\bf r}\right)
\exp\left\{-2\kappa\int_0^t d\tau\left({\bf k}\hat{W}^{-1}(\tau)
\hat{W}^{-1,T}(\tau){\bf k} \right)\right\}\Bigr\rangle.
\label{pco1}
\end{eqnarray}
We study the large time ${\cal G}t \gg 1$ asymptotics of this tensor. The key point which simplifies all the calculations in the dynamo problem 
lies in the fact that in this limit the averaging  of products $B_\alpha(\bm{r},t)B_\mu(\bm{r}',t)\dots $ with respect to the measure (\ref{mts}) can be done in saddle-point approximation at least up to a constant factor.  This means that the dominating contribution to the 
required expectation values are given by trajectories $\rho(t)=A{\cal G}t)+\delta\rho(t)$ where $A$ is some positive constant and 
$\delta\rho(t)\sim 1 \ll {\cal G}t$. This property is a consequence  of proportionality of the product $B_\alpha(\bm{r},t)B_\mu(\bm{r}',t)\dots $
to $\exp\left(a\rho(t)\right)$ with some positive constatnt $a>0$. Substituting $\rho(t)=A{\cal G}t)+\delta\rho(t)$ into the path integral for 
 $\langle B_\alpha(\bm{r},t)B_\mu(\bm{r}',t)\dots \rangle$ one can find the constant $A(a)$ requiring the absence of the terms leading to linear growth of $\delta\rho(t)$ with $t$. The exponential growth of $\exp(\rho(t))$ on the dominating trajectories leads to suppression of  $\dot{\chi}^2$ for ${\cal G}t\gg 1$ \cite{CFK,CKV}. The fluctuations of the variable $\chi$ become frozen  in sense that the probability distribution function (PFD) of $\chi$ is asymptotically time-independent for $\rho(t)=A{\cal G}t)+\delta\rho(t)$. This statement can be illustrated by the joint PDF for the variables $(\rho, \chi, \varphi)$ found in  \cite{K10}. For  ${\cal G}t\gg 1$ it has the form:
\begin{eqnarray}
{\cal P}(\rho,\chi,\varphi,t)\approx
{\cal C}\frac{\rho}{({\cal G}t)^{3/2}}
\left(1+\chi^2\right)^{-\frac{1}{2}-\frac{\rho}{2{\cal G}t}}
\exp\left[-\frac{\rho^2}{2{\cal G}t}+\rho-\frac{1}{2}{\cal G}t\right],\quad  {\cal C}\sim 1.
\label{assy}
 \end{eqnarray} 
Here some inaccuracies made in \cite{K10} are corrected. The independence of (\ref{assy}) on $\varphi$ corresponds to the uniform distribution 
of this angle in the interval $[0,2\pi]$. 
 
We consider here the limit of small dissipation parameter $\kappa$ so that the corresponding dissipation length 
$r_d=2\sqrt{\kappa/{\cal G}}$ is small comparing with the initial correlation length $l$: $r_d\ll l$. On the other hand, the dissipation 
becomes important when time goes to infinity and it determines to a large extent the amplitude and the form of the field correlation functions.

Returning to the expression (\ref{pco1}) and taking into account the independence of the parameter $\chi$ on time at ${\cal G}t\gg 1$ we perform 
the change of variables 
\[
{\bf k}=\left(\begin{array}{cc} 1 & 0 \\
        -\chi  & 1  \end{array}\right) {\bf Q}        
\]
and neglect  the decaying exponential $\exp(-\rho)$ everywhere.
For dominating trajectories the approximation 
\begin{equation}
\label{limt}
\int\limits_0^t d\tau\, e^{2\rho(\tau)} 
\approx c {\cal G}^{-1}e^{2\rho(t)}. 
\end{equation}
is valid.
Here the quantity $c$: 
\[
c={\cal G}\int\limits_0^t d\tau\, e^{2\rho(\tau)-2\rho(t)}
\]
fluctuates but such the fluctuations result in some numerical factor when we compute the correlator (\ref{pco1}). This effect 
can be important in study of high-order moments of the magnetic field but it is not important when we compute $F_{\alpha\beta}$ 
 up to a constant prefactor. Hence the leading comtribution to the correlation tensor has the form:
\begin{eqnarray}
&&F_{\alpha\beta}({\bf r},t)\propto\Bigl\langle (1+\chi^2)e^{2\rho}
\left(\begin{array}{cc} \cos^2\varphi & -\sin \varphi\cos \varphi \\
        -\sin\varphi\cos\varphi  & \sin^2\varphi  \end{array}\right)_{\alpha\beta}\times
\nonumber \\
&&\times\int\,d^2{\bf Q}f\left(Q_1^2l^2[1+\chi^2]\right)
\exp\left\{iQ_2e^\rho\left(r_1\sin\varphi+r_2\cos\varphi\right)-c Q_2^2e^{2\rho}r_d^2\right\}
\Bigr\rangle.
\label{pco2}
\end{eqnarray}
Here  the averaging is assumed to be performed with respect to the finite-dimensional probability density (\ref{assy}). $dQ_1$ - integration can done  first producing the factor $\sim (1+\chi^2)^{-1/2}l^{-1}f_0$, $f_0=\int dq\,f(q^2)$. The subsequent integration 
with respect to $d\chi$ is convergent and results into a numerical factor of the order of unity. The integration over the variable $Q_2$ is Gaussian and it gives us the expression:
\begin{equation}
F_{\alpha\beta}({\bf r},t)\propto\frac{f_0}{lr_d}\Bigl \langle e^{\rho}
\left(\begin{array}{cc} \cos^2\varphi & -\sin \varphi\cos \varphi \\
        -\sin\varphi\cos\varphi  & \sin^2\varphi  \end{array}\right)_{\alpha\beta}
\exp\left[-\frac{r^2}{c r_d^2}\sin^2\left(\varphi+\varphi_0\right)\right]
\Bigr\rangle.
\label{pco21}
\end{equation}
where $\tan\varphi_0=r_2/r_1$. The averaging over the variable $\rho$ produces growing factor $\exp(3{\cal G}t/2)$ which corresponds to the dominating trajectory $\rho(t)\approx 2{\cal G}t$. The final averaging over the angle $\varphi$ results in universal forms for correlation tensor in two limits:
\begin{eqnarray}
\label{te1}
F_{\alpha\beta}({\bf r},t)\propto \delta_{\alpha\beta}\frac{f_0e^{3{\cal G}t/2}}{l r_d},
\quad r\ll r_d,
\end{eqnarray} 
and 
\begin{equation}
\label{te4}
F_{\alpha\beta}({\bf r},t)\propto \frac{f_0}{l }\exp\left(\frac{3}{2}{\cal G}t\right)
\frac{r_\alpha r_\beta}{r^3}, \quad r\gg r_d.
\end{equation} 
In this asymptotics the correlation tensor $F_{\alpha\beta}({\bf r},t)$  does not depend on $r_d$ similar to three-dimensional case \cite{Kaz}. 

Turning to the four-point correlation tensor it is easy to see that for the linear velocity profile (\ref{linprof}) there is the decomposition:
\begin{equation}
\label{four1}
\langle B_\alpha({\bm r}_1,t) B_\beta({\bm r}_2,t)B_\gamma({\bm r}_3,t)B_\mu({\bm r}_4,t)\rangle=
\hat{Q}\left({\bm R}_{12}, {\bm R}_{34}\right)+ \hat{Q}\left({\bm R}_{13}, {\bm R}_{24}\right)+
\hat{Q}\left({\bm R}_{14}, {\bm R}_{23}\right),
\end{equation} 
where ${\bm R}_{jl}={\bm r}_j-{\bm r}_l,\, j,l=1,\dots,4$. The tensor function $\hat{Q}\left({\bm R}_1, {\bm R}_2\right)$ can be expressed in a form similar to (\ref{pco21}). 
The presence of two angles related to the vectors ${\bm R}_1$ and ${\bm R}_2$ leads to the strong angle dependence:
\begin{equation}
\label{angl}
\hat{Q}\left({\bm R}_1, {\bm R}_2\right)\sim \exp\left(-\frac{R^2}{r_d^2} \Theta_{12}^2\right),
\end{equation}  
where $ \Theta_{12}$ is the angle between ${\bm R}_1$ and ${\bm R}_2$ and the relation $R_1\sim R_2\sim R$ is assumed. In the collinear limit ${\bm R}_{1,2}={\bm n}R_{1,2}$ 
one can get the simple expression:
\begin{equation}
\label{coll}
\hat{Q}\left({\bm R}_1, {\bm R}_2\right)\sim n_\alpha n_\beta n_\gamma n_\mu \frac{f_0^2 l^2}{r_d\sqrt{R_1^2+R_2^2}}\exp\left(4{\cal G}t\right).
\end{equation}

\section{Conclusion.}

The picture corresponding to the correlators found above is the following: the field $B_\alpha$ is concentrated in parallel strips of the width $\sim r_d$. 
The field is directed mainly along the strip. These strips are rotated by the flow and the time interval $\tau_s$ when the observation 
points reside in a strip is of the order of $r_d/r$ where $r$ is the distance between the points. The angle dependence (\ref{angl}) corresponds exactly to this 
intermittent stripes structure.
The $r$-dependence of the two- and four-point correlators coincides because it has the mentioned geometrical origin. 
The ratio $\langle {\bm B}^4\rangle/\langle {\bm B}^2\rangle^2\sim \exp(2{\cal G}t)\gg 1$ corresponds also to the temporal intermittency.

For the times $t\gg {\cal G}^{-1}\ln(R_c/r_d)$ the approximation (\ref{linprof}) does not work. The pair correlator of the magnetic field is studied in \cite{Kolo16} but 
high-order correlation functions are unknown and the question about the intermittency effects in this stage of evolution remains open.

I am very grateful to V.V.Lebedev for helpful advices and numerous discussions and to L.Krainov for useful remarks. I wish to thank G.Falkovich for the interest to the work.
The author acknowledges support of RScF, grant 14-22-00259.


\section*{References}
 \begin{thebibliography}{99}

 \bibitem{MHD}
 H. K. Moffatt,  Magnetic Field Generation in Electrically Conducting Fluids. 
 Cambridge University Press, 1978.

 \bibitem{P79}
 E. N. Parker,  Cosmic magnetic fields their origin and activity. 
 Clarendon Press Oxford, 1979.

 \bibitem{ZRS}
 Ya. B. Zeldovich,   A. A. Ruzmaikin,  D. D. Sokolov,  
 The Almighty Chance.  World Scientific, 1990.

 \bibitem{astro1}
 G. R\"uediger  and R. Hollerbach,  
 The Magnetic Universe: Geophysical and Astrophysical Dynamo Theory. 
 Wiley, 2004.

 \bibitem{schek}
 A. A. Schekochihin, S.C. Cowley, S. F. Taylor, L. Maron,  J. C. McWilliams, 
 Simulations of the small-scale turbulent dynamo.
 Astrophysical Journal {\bf 612} 276-307, (2004).


 \bibitem{00GS}
A. Groisman, V. Steinberg, 
 Elastic Turbulence in a Polymer Solution Flow.
 Nature,  {\bf 405} 53 (2000). 

 

 \bibitem{01GSb}
 A. Groisman, V. Steinberg, 
 Efficient Mixing at Low Reynolds Numbers Using Polymer Additives.
 Nature, {\bf 410} 905 (2001).
 
 \bibitem{00BFL}
 E. Balkovsky, A. Fouxon A., V. Lebedev, 
 Turbulent Dynamics of Polymer Solutions.
 Phys. Rev. Lett. {\bf 84} 4765-4768 (2000).
 
  \bibitem{LL}
 L. D. Landau, E. M. Lifshitz, 
 Course of theoretical physics.
 Electrodynamics of Continuous Media. Pergamon, 1984.
 
 \bibitem{Kaz}
 A. P. Kazantsev,
 Enhancement of a magnetic field by a conducting fluid. 
 Sov. Phys JETP, {\bf 26} 1031 (1968).

 \bibitem{99CFKV}
 M. Chertkov, G. Falkovich, I. Kolokolov, M. Vergassola, 
 Small-scale Turbulent Dynamo.
 Phys. Rev. Lett., {\bf 83} 4065-4068 (1999).
 
 \bibitem{01FGV}
 G. Falkovich, K. Gawedzki, M. Vergassola, 2001
 Particles  fields in fluid turbulence.
 Rev. Mod. Phys. {\bf 73} 913.
 
\bibitem{KleSiz} 
I.V. Kolokolov, V.V. Lebedev, G.A.Sizov,  Magnetic field correlations in a random flow with
strong steady shear,  ZhETF, {\bf 140}, 387,(2011)  

\bibitem{Zeldovich1}
 Ya. B. Zeldovich,
 The magnetic field in the two-dimensional motion of a conducting turbulent fluid,
 ZhETF {\bf 31} 154 (1956)
 [Sov. Phys. JETP {\bf 4} 460 (1957)].

 \bibitem{Zeldovich2}
 Zeldovich Ya. B. and Ruzmaikin A. A.,
 Magnetic field of a conducting fluid in two-dimensional motion,
 ZhETF {\bf 78} 980 (1980) [Sov. Phys. JETP {\bf 51} 493 (1980)].

 \bibitem{Zeldovich3}
 Ya. B. Zeldovich, A. A. Ruzmaikin, S. A. Molchanov and  D. D. Sokolov,
 Kinematic Dynamo Problem in a Linear Velocity Field,
 J. Fluid Mech. {\bf 144} 1 (1984).

  \bibitem{KogKolLe} 
 V.R. Kogan, I.V. Kolokolov and V.V. Lebedev, Kinematic magnetic dynamo in a random flow with
strong average shear, J. Phys. A: Math. Theor. {\bf 43}, 182001,(2010)  

\bibitem{Kolo16} I.V. Kolokolov , Evolution of magnetic field fluctuations in two-dimensional chaotic flow.
arXiv: 1603.08771v1, 2016.

\bibitem{Ose}
V.Oseledets, Geophysical and Astrophysical Fluid Dynamics, {\bf 73}, 133 (1993).
\bibitem{Arno}
V.I.Arnold, B.A.Khesin, Topological Methods in Hydrodynamics, Springer-Verlag New York, Inc., 1998.

\bibitem{Sche2} A.A.Schekochihin, S.C.Cowley, S.F.Taylor, G.W.Hammett, J.L.Maron, and J.C.McWilliams,
Saturated State of Nonlinear Small-Scale Dynamo, 
 Phys. Rev. Lett. {\bf 92}, 084504 (2004).

\bibitem{ChGK}M.~Chertkov, A.~Gamba, I.Kolokolov, Exact field-theoretical description of passive scalar
convection in an N-dimensional long-range velocity field,
Phys. Lett. A {\bf 192}, 435-443 (1994). 

 \bibitem{CFK}
M.Chertkov, G.Falkovich, I.Kolokolov, Intermittent dissipation of a passive scalar in turbulence,
 Physical Review Letters, {\bf 80} , p.2121 (1998).

 \bibitem{Kam}
Alex Kamenev, Field Theory of Non-Equlibrium Systems, Cambridge Univ. Press, 2011. 

 \bibitem{CKV}
M. Chertkov, I.Kolokolov, M. Vergassola, Inverse cascade and intermittency of passive scalar in 1d smooth flow,
{\it Physical Review} E, {\bf 56},  pp.5483-5499 (1997) 

 \bibitem{K10}
 I.V. Kolokolov, 
Statistical Geometry of Chaotic Two-Dimensional Transport, JETP Lett. {\bf 92}, 107 (2010).

\end{thebibliography}
\end{document}